\documentstyle[12pt]{article}
\begin{document}
\newcommand{\be}{\begin{equation}}
\newcommand{\ee}{\end{equation}}

\begin{center}

{\bf Cherenkov radiation by particles\\
traversing the background radiation}\\

\vspace{3mm}

I.M. Dremin\\

Lebedev Physical Institute, Moscow 119991, Russia\\ 

\end{center}

\begin{abstract}
High energy particles traversing the Universe through the cosmic microwave
backgroung radiation can, in principle, emit Cherenkov radiation. It is shown
that the energy threshold for this radiation is extremely high and its
intensity would be too low due to the low density of the "relic photons gas"
and very weak interaction of two photons.

\end{abstract}

Recently, the particles with energies exceeding $10^{20}$ eV were observed
in cosmic ray studies. Their sources have not yet been identified, but they
originate most likely outside of our Galaxy. One could hope that some knowledge
about the distance from these sources would be possible to obtain, in principle,
from studies of Cherenkov radiation by such particles since its intensity
is proportional to the distance covered by a particle.

Let us consider Cherenkov radiation by a high energy particle traversing
the "gas" of relic photons with the temperature $2.73^0$ K.
Even though the density $\nu $ of such photons is very low in the Universe
$\nu \approx 500$ photons/cm$^3$, the particle path could be rather large
up to tens Mpc (1 Mpc $\approx 3\cdot 10^{24}$ cm), and one could hope to
register its
Cherenkov radiation because the intensity is proportional to the path length.

The necessary conditions for Cherenkov radiation to be observed are the excess
of the index of refraction $n$ over 1, i.e.
\be
\Delta n=n-1>0      \label{delt}
\ee
and the real emission angle, given by the formula 
\be
\cos \theta=\frac {1}{\beta n},     \label{cost}
\ee
where $\beta =v/c=\sqrt {1-\frac {m^2}{E^2}},  m, E$ are the particle mass and
energy. For small values of $m/E$ and $\Delta n$ one gets
\be
\theta \approx \sqrt {2\Delta n - \frac {m^2}{E^2}}.  \label{thet}
\ee
Herefrom, the condition for the energy threshold $E_{th}$ is written as
\be
E\geq E_{th}=\frac {m}{\sqrt {2\Delta n}}.     \label{ethr}
\ee
It is easily seen that the threshold can become very high for small $\Delta n$.

The number of Cherenkov photons emitted by a particle with the electric charge
$e$ in the interval of frequencies $d\omega $ from the path length $dl$ is
given by the common expression \cite{tfra}
\be
\frac {dN}{d\omega dl}=2\alpha \Delta n,     \label{adn}
\ee
where the fine structure constant $\alpha =e^2\approx 1/137$.
Thus all physical characteristics of the process are determined by the value
$\Delta n$. The intensity of the radiation (\ref{adn}) decreases with the
threshold energy (\ref{ethr}) increase:
\be
\frac {dN}{d\omega dl}=\frac {\alpha m^2}{E_{th}^2}.     \label{adn1}
\ee

Surely, it is possible to use the notion of the medium (and, consequently,
the macroscopic approach) only in 
the case of extremely long-wave radiation for the very diluted gas of relic
photons. At the same time, it is well known that in usual media the value
of $\Delta n$ is uniquely related to the polarization operator of the medium.
If this relation is valid also in the case treated here, one must consider
the polarization operator of the light-light scattering or the real part of the
forward elastic light-light scattering amplitude ${\rm Re} F(\omega , 0^0)$.
For the index of refraction slightly different from one\footnote{The more general
formula of Lorentz-Lorentz (see \cite{dwat}, p. 693) can be applied without this
restriction.}, this relationship is given in the quantum scattering theory
by the common formula \cite{dwat, drem}:
\be
\Delta n=\frac {2\pi \nu }{\omega ^2}{\rm Re} F(\omega , 0^0) ,   \label{deln}
\ee
where the elastic scattering amplitude of the two gamma-quanta $F(\omega )$
has been normalized to the total cross section
$\sigma _{\gamma \gamma }(\omega )$ according to the optical theorem
\be
{\rm Im} F(\omega , 0^0)=\frac {\omega }{4\pi }\sigma _{\gamma \gamma }(\omega ).    \label{opti}
\ee
Using these formulas, one easily gets
\be
\Delta n=\frac {\nu \sigma _{\gamma \gamma }\rho }{2\omega },    \label{den1}
\ee
where $\rho ={\rm Re} F(\omega , 0^0)/{\rm Im} F(\omega , 0^0)$.
Herefrom and from (\ref{adn}), it follows that
\be
\frac {dN}{d(\ln \omega )dl} = \alpha \nu \sigma _{\gamma \gamma } \rho, \label{dndd}
\ee
and
\be
E_{th}=\frac {m\sqrt {\omega }}{\sqrt {\nu \sigma _{\gamma \gamma }\rho }} ,    \label{etfr}
\ee
i.e. the radiation threshold increases with the energy of the registered quanta.

The impressive strong limit is imposed on the energy threshold for
Cherenkov radiation in the gas of relic photons. In electrodynamical processes,
the total cross section $\sigma _{\gamma \gamma }$ reaches its maximum value of
about $1.6 \mu b$ at the total energy of the two photons in their center
of mass system $\omega _c \approx 3m_e$ where the electron mass is
$m_e\approx 0.5$ MeV (see \cite{kneu, deto}). At lower energies, below the
threshold for the creation of the electron-positron pair (in particular, for
a visible light), only elastic scattering is important, the real part of the
amplitude dominates and the cross section decreases at $\omega \rightarrow 0$
proportionally to $\omega _c^6$ (see the discussion at the end of the paper).
At high energies, as seen from the formula (\ref{den1}), the value of $\Delta n$ 
tends to zero at $\omega \rightarrow \infty $, and, therefore, the decline
from the linear dispersion law $\omega =k$ can become noticeable only in the
region near the maximum of the cross section $\sigma _{\gamma \gamma }$.

The c.m.s. energy of the two-photon system varies from 0 (when photons move in
the same direction in the laboratory system) to its maximum value
\be
\omega _c=2\sqrt {\omega \omega _r},   \label{ooo}
\ee
when the relic photon with energy $\omega _r$ moves in the laboratory system
in the opposite direction
to the emitted quantum with energy $\omega $. For the relic photons
$\omega _r\approx 2.4\cdot 10^{-4}$ eV. Thus the energy $\omega _c\approx 3m_e$
can be only achieved if the particle emits quanta with the energy
\be
\omega =\frac {\omega _c^2}{4\omega _r}\approx 2\cdot 10^{15} {\rm eV}. 
\ee

First, let us consider the high energy quanta which can create the 
electron-positron pairs in the cosmic microwave background radiation.
In this case of very energetic radiation quanta $\Delta n <10^{-48}$, if one
inserts high-energy values of
$\rho \sim 0.1$. Then for the proton with mass $m$=1 GeV one gets an estimate
\be
E_{th}>10^{33} {\rm eV}.      \label{egre}
\ee
The particles with such high energy have not yet been observed anywhere.
Such a low index of refraction and, correspondingly, high value of the energy
threshold
are determined by the low density of relic photons in the Universe
and by the small total cross section of the photon-photon interaction.

The estimate of the upper limit of the intensity of the high-energy Cherenkov
radiation at the path length $L$
\be
N<\alpha \nu \sigma _{\gamma \gamma }L     \label{nans}
\ee
is obtained from the formula (\ref{dndd}), if one takes into account that
$\Delta \omega /\omega <1$ and $\rho <1$. If the path, wherefrom the radiation
is collected, is equal to $L\sim $ 1 Mpc $\sim 3\cdot 10^{24}$ cm, and the
maximum value of $\sigma _{\gamma \gamma }$ is about 1.6 $\mu b$
(see \cite{kneu, deto}) then for $\nu \approx 500$ photons/cm$^3$ the following upper
limit on the number of emitted quanta is imposed according to (\ref{nans}) 
\be
N<3\cdot 10^{-5}.     \label{nles}
\ee

Thus, the intensity of the high-energy Cherenkov radiation in the gas of relic
photons is too low to be observable even if a proton passes through the Universe
along hundreds Mpc. For a primary nucleus, one should insert in the formula
(\ref{adn}) its total charge, and, therefore, the intensity
of radiation increases in proportion to the nucleus charge squared. However,
the total threshold energy increases proportionally to its mass number so that
the threshold is much higher for nuclei compared to protons. At the same time,
let us note that according to formula (\ref{ethr}) the threshold energy per
nucleon of a nucleus is only determined by the value of $\Delta n$, and,
therefore, it is the same for all nuclei.

Some contribution to the polarization operator of the light-light scattering
is, in principle, provided by the hadronic component of the photon structure
function as well. It can be accounted in the framework of the vector dominance
model when the quanta are transformed into virtual $\rho $-mesons which
interact resonancely. The real part of the scattering amplitude is with
necessity positive in one of the halves of the resonance peak as clearly follows
from the Breit-Wigner formula. The cross section
$\sigma _{\gamma \gamma }$ slightly increases. Nevertheless, this does not
lead either to the diminished threshold value in the formula (\ref{egre}),
or to the increased intensity of Cherenkov radiation because of the increase
of $\omega $ due to $\omega _c$ increasing up to the values of the order 
of the $\rho $-meson mass.
The hadronic contribution to the ratio
$\rho ={\rm Re} F/{\rm Im} F$ can be also positive at higher energies as it is
well known from experiments in all studied hadronic reactions in complete
accordance with predictions of dispersion relations. However, the threshold
of the radiation again increases in this case due to the further increase of the required
values of $\omega _c$.

The estimates have shown that the derived above
conclusions are "protected" by several orders of magnitude as is already seen
from the formulas (\ref{egre}) and (\ref{nles}), and therefore they are robust.

Let us estimate the absorption of Cherenkov radiation at cosmic distances.
For a plane wave $e^{ikr}$ it is given by a factor
\be
\exp [-\omega L {\rm Im} n(\omega )]\approx \exp [-\nu \sigma _{\gamma \gamma }
L/2],   \label{abso}
\ee
wherefrom the absorption length $L_{abs}$ is estimated as
\be 
L_{abs}=\frac {2}{\nu \sigma _{\gamma \gamma }}>2\cdot10^{27} {\rm cm}.   \label{labs}
\ee
The absorption length of the gamma-quantum in the relic radiation gas depends
on the energy of the gamma-quantum only due to the cross section energy
dependence and it is so large that the damping of Cherenkov radiation can be
neglected for the distances shorter than thousands Mpc. Correspondingly,
the energy losses which determine the red shift are very small at such
"short" distances $L\ll L_{abs}$:
\be
\frac {\Delta \omega }{\omega }=1-\exp (-\frac {\nu \sigma _{\gamma \gamma }L}
{2})\approx \frac {\nu \sigma _{\gamma \gamma }L}{2}\ll 1.
\ee

Thus, the final conclusion is that the high-energy Cherenkov radiation by
particles traversing
the gas of relic photons is impossible to observe, first of all, because
until now there have been detected no particles with such high energies in
the Nature exceeding the required energy threshold. Even if such particles
were registered, the intensity of the high-energy Cherenkov radiation would,
possibly, be too low to detect it.

The principal possibility of the Cherenkov radiation
by a charged particle in the intergalactic space is, however, not excluded if
there exists "an intergalactic medium component" with higher density and
larger cross section of interaction with photons.

Another possibility is related to studies of quanta with energies lower
than the threshold for creation of the electron-positron pair, i.e. for
$\omega <10^{15}$ eV. Even though
the cross section decreases, the ratio $\rho $ becomes very large. To estimate 
the index of refraction for such quanta, the classical consideration
with Maxwell equations and dispersion relations \cite{mais}
or the effective lagrangian with 4-photon interaction \cite{thom}, describing
the deviation from the classical theory by quantum effects, have been used.
More "optimistic" estimates of $\Delta n\sim 10^{-41}$ follow, e.g., from Fig. 2
of Ref. \cite{mais}, where, in our notations, the variables $k$ and $z$ read
\be
k=m_{e}^4/16\pi ^2\alpha ^2\nu \omega _r \approx 8\cdot 10^{40}; \;\;\;\;
z=m_{e}^2/\omega \omega _r = 4m_{e}^2/\omega _c^2.
\ee
This increase of $\Delta n$ is
because of much larger values of $\rho $. This would lead to
lower threshold $E_{th}> 10^{29}$ eV. Nevertheless, this value of the
threshold energy is still
very high. The intensity of radiation could, however, become quite noticeable
$N\sim 10$. In Ref. \cite{thom} the value of $\Delta n$ at low energies was
estimated as $4.7\cdot 10^{-43}$.
Let us remind that the notion of the medium in classical approach
can be used when the wavelength of the impinging photon is larger than
typical distances inside the medium.

At the same time, $\Delta n$ increases \cite{thom} with the temperature $T$ in
equilibrium as $T^4$. Therefore, at the early stage of the Universe, when the
cosmic microwave background radiation separated from the matter
at the temperature about $3000^0$K, the index of refraction was
$\Delta n\sim 10^{-30}$, i.e. the threshold energy was $E_{th}\sim 10^{24}$ eV.
This estimate is closer to "realistic" but still rather high energies. 

Thus the final conclusion about extremely high energy threshold for Cherenkov
radiation in the cosmic microwave background radiation is left intact.

I am grateful to L.G. Tkatchev who pointed out this problem to me and to
I.V. Andreev, B.M. Bolotovsky, A.D. Erlykin, E.L. Feinberg, I.F. Ginzburg,
V.L. Ginzburg and V.A. Maisheev for discussions and comments.

This work is supported by the RFBR grant 00-02-16101.

\end{document}